\documentclass[amsmath,amssymb,prl,twocolumn]{revtex4}

\usepackage{graphicx}

\def\de{\delta}

\def\et{\eta}

\def\rh{\rho}

\def\si{\sigma}

\def\ph{\phi}

\def\ch{\chi}

\def\om{\omega}

\def\lsim{\mathrel{\rlap{\lower4pt\hbox{\hskip1pt$\sim$}}
    \raise1pt\hbox{$<$}}}
\def\gsim{\mathrel{\rlap{\lower4pt\hbox{\hskip1pt$\sim$}}
    \raise1pt\hbox{$>$}}}
\def\sqr#1#2{{\vcenter{\vbox{\hrule height.#2pt
         \hbox{\vrule width.#2pt height#1pt \kern#1pt
         \vrule width.#2pt}
         \hrule height.#2pt}}}}
\newcommand{\beq}{\begin{equation}}
\newcommand{\eeq}{\end{equation}}
\newcommand{\bea}{\begin{eqnarray}}
\newcommand{\eea}{\end{eqnarray}}

\def\etal{{\it et al.}}
\def\re{{\rm Re}}
\def\im{{\rm Im}}

\def\voc{\mathrel{\rlap{\lower0pt\hbox{\hskip1pt{$c$}}}
    \raise3pt\hbox{$\neg$}}}
\def\sk#1#2#3{#1^{(#2)}_{#3}}
\def\cjm#1#2#3{c^{(#1)}_{(#2)#3}}

\def\cIdjm#1#2{\cjm{#1}{I}{#2}}
\def\cf{\sk{(\voc_F^{(d)})}{0E}{njm}}
\def\cfdnjm#1#2{\sk{(\voc_F^{(#1)})}{0E}{#2}}

\begin{document}

\title{Cavity Bounds on Higher-Order Lorentz-Violating Coefficients}

\author{Stephen R. Parker,$^1$ Matthew Mewes,$^2$ Paul L. Stanwix,$^1$ and Michael E. Tobar$^1$} 
\affiliation{
$^1$School of Physics, The University of Western Australia, Crawley WA 6009, Australia\\
$^2$Department of Physics and Astronomy, Swarthmore College, Swarthmore, Pennsylvania 19081, USA}

\date{\today}

\begin{abstract}
We determine the sensitivity of a modern
Michelson-Morley resonant-cavity experiment
to higher-order nonbirefringent and
nondispersive coefficients of the Lorentz-violating
Standard-Model Extension.
Data from a recent year-long run of the experiment
is used to place the first experimental bounds on
coefficients associated with nonrenormalizable
Lorentz-violating operators.
\end{abstract}

\maketitle

Over the past century,
the behavior of light has been
scrutinized for possible defects in
Einstein's theory of special relativity.
Recent advances in theory and experiment
have led to renewed interest and
a large number of tests of Lorentz invariance,
the symmetry behind relativity \cite{datatables}.
These studies are motivated, in part,
by the possibility that theories incorporating 
quantum gravity may lead to small violations
of Lorentz invariance at low energies,
providing an avenue for searches
for fundamental physics \cite{ks}.
New work has uncovered
a very large class of unexplored 
higher-order violations in photons \cite{km09}.
This letter provides the first laboratory-based 
constraints on these violations,
initiating a new class of tests of
Lorentz invariance.

A wide variety of experimental techniques
have been employed in searches for
Lorentz violation \cite{datatables}.
One particularly sensitive 
type of experiment is based on
electromagnetic resonant cavities \cite{cav1,cav2,micro,optical,cav3}.
These are considered descendents of the classic
Michelson-Morley experiment \cite{mm} 
and are analyzed using a theoretical framework known
as the Standard-Model Extension (SME) \cite{ck,akgrav}.
While the SME describes general violations of
Lorentz invariance in any system,
most studies have focused on the minimal SME,
which restricts attention to operators of
renormalizable dimension in flat spacetime.
However, the photon sector of the SME has
recently been extended to include
operators of arbitrary dimension \cite{km09}.
Here we search for the effects of these
higher-order nonrenormalizable violations
using the cavity-based experiment described
in \cite{cav1}.

The photon sector of the nonminimal SME
contains a large number of possible violations
that lead to different unconventional phenomena.
It is convenient to split these violations
into subclasses based on their physical effects.
Two subclasses that are particularly useful
are based on cosmological birefringence and dispersion.
These unconventional features affect the way
light propagates through empty space and
can be tested with extreme precision
using astrophysical sources \cite{bire,km_apjl,fermi}.
Violations that do not cause birefringence or
dispersion are not easily detected in 
astrophysical observations.
However, they can often be measured
in laboratory-based experiments,
which complement astrophysical tests.

In the SME, the higher-order
nonbirefringent nondispersive violations
are controlled by a set of coefficients
that are denoted by $\cf$.
These are referred to as camouflage coefficients,
due to their benign nature, and
are the primary focus of this work.
The $d$ index on the coefficients is the
dimension of the associated operator
and only takes on even values $d\geq 6$.
It should be noted that there are
nonbirefringent nondispersive violations
among the renormalizable $d=4$ operators.
The coefficients associated with these
operators are $\cIdjm{4}{jm}$ and were
the main focus of previous 
cavity experiments \cite{cav1,cav2,micro,optical,cav3}.
We will include the renormalizable
coefficients in our analysis for completeness.
We will restrict our analysis of nonrenormalizable
violations to the two lowest-order cases,
$d=6,8$, for simplicity.
However, the techniques outlined here could be
applied to higher-order violations.

The Lorentz violations associated with
the $\cIdjm{4}{jm}$ and the $\cf$
coefficients lead to tiny shifts in
the resonant frequencies of cavities.
As described in Ref.\ \cite{km09},
the frequency shift resulting
from coefficients of a given
dimension takes the generic form
\beq
\frac{\de\nu}{\nu} = \sum_{njm} {\cal M}^{\rm cav}_{njm} c^{\rm cav}_{njm} \ ,
\eeq
to first order in coefficients for Lorentz violation.
The index $n$ characterizes the frequency/wavelength
dependence of the coefficient,
but is irrelevant for dimension $d = 4$.
The $jm$ indices represent the usual
angular-momentum quantum numbers,
which characterize the rotational 
properties of the coefficients.
Note that a violation of rotation
invariance serves as our signal of
Lorentz violation.

The ${\cal M}^{\rm cav}_{njm}$
are experiment-dependent constants that
determine the sensitivity to each coefficient
and are calculated using the conventional
solutions to the Maxwell equations in the cavity.
The ``${\rm cav}$'' designation indicates a reference
frame fixed to the cavity with the $z$ direction
along the symmetry axis.
This frame is not inertial,
so the $c^{\rm cav}_{njm}$ 
coefficients change as the cavity rotates.
We need to express the frequency in terms of
fixed coefficients in an inertial frame,
which by convention is taken as a Sun-centered
celestial equatorial frame \cite{datatables}.
Ignoring boosts,
the transformation is achieved by a series
of rotations that, in the current language,
take the form of Wigner matrices.
The result can be written 
$\de\nu/\nu = \sum_{mm'} A_{mm'} e^{im\ph + im'\om_\oplus T}$,
where
\beq
A_{mm'} = \sum_{nj}
{\cal M}^{\rm cav}_{njm'}
d^{(j)}_{m'm}(-\tfrac{\pi}{2})
d^{(j)}_{mm'}(-\ch)
c^{\rm Sun}_{njm'} \ .
\label{eq:Amm}
\eeq
In these expressions,
$d^{(j)}_{mm'}$ are the
``little'' Wigner matrices,
$\ch$ is the colatitude of the laboratory,
$\om_\oplus T$ is the
right ascension of the laboratory,
and $\ph$ is the angle between south and
the cavity symmetry axis.
The $c^{\rm Sun}_{njm'}$ are constant
Sun-frame coefficients.
Here we assume the cavity lies in the
horizontal plane.
The $A_{mm'}$ amplitudes obey the relation
$A^*_{mm'} = A_{(-m)(-m')}$.
This implies that the frequency shift is real,
as expected.

The experiment involves monitoring the beat
frequency between resonances of two identical
horizontal rotating cavities.
Violations of rotation symmetry can introduce
orientation dependence.
This would manifest as a dependence on the
orientation with respect to the laboratory ($\ph$)
or in variations at the sidereal frequency
$\om_\oplus \simeq 2\pi/(\mbox{23 hr 56 min})$
as the laboratory rotates throughout the day.
Since the cavities of interest are perpendicular,
we let one cavity lie at an angle $\ph$
and the other at an angle $\ph+\pi/2$.
This leads to 
$\nu_{\rm beat}/\nu = \sum A^{\rm beat}_{mm'} e^{im\ph + im'\om_\oplus T}$,
where
\beq
A^{\rm beat}_{mm'} = (1-i^m) A_{mm'}\ .
\eeq
The beat amplitudes
$A^{\rm beat}_{mm'}$ obey the same
conjugation relationship as the $A_{mm'}$
amplitudes.

It is convenient to express the beat frequency
in terms of sines and cosines with real amplitudes.
Trigonometric expansions can then be used to mix
together frequency sidebands of the same multiple
of sidereal frequency, $m'$,
which has the benefit of combining the
positive and negative $m$ and $m'$
terms allowing the data to be fit to all components.
This gives
\beq
\frac{\nu_{\rm beat}}{\nu} = \sum_{m\geq 0} 
\big[ C_m(T) \cos (m\ph) + S_m(T)\sin (m\ph) \big] \ ,
\label{eq:nubeattwicerot}
\eeq
where $C_m(T)$ and $S_m(T)$ are amplitudes that vary slowly,
at integer multiples of the sidereal frequency.
They are given by
\begin{align}
  C_m(T) &= \sum_{m'\geq 0}\!
  \big[C^C_{mm'} \cos (m'\om_\oplus T) 
    + C^S_{mm'} \sin (m'\om_\oplus T) \big]\, ,
  \notag \\
  S_m(T) &= \sum_{m'\geq 0}\!
  \big[ S^C_{mm'} \cos (m'\om_\oplus T) 
    + S^S_{mm'} \sin (m'\om_\oplus T) \big]\, ,
\end{align}
with
\begin{align}
  C^C_{mm'} &= 2\, \et_m \et_{m'}\, \re\big[A^{\rm beat}_{mm'} + A^{\rm beat}_{m(-m')}\big] \ ,
  \notag \\
  C^S_{mm'} &= -2\, \et_m\, \im\big[A^{\rm beat}_{mm'} - A^{\rm beat}_{m(-m')}\big] \ ,
  \notag \\
  S^C_{mm'} &= -2\, \et_{m'}\, \im\big[A^{\rm beat}_{mm'} + A^{\rm beat}_{m(-m')}\big] \ ,
  \notag \\
  S^S_{mm'} &= -2\, \re\big[A^{\rm beat}_{mm'} - A^{\rm beat}_{m(-m')}\big] \ ,
  \label{eq:CandScomps}
\end{align}
where, for convenience,
we define $\et_0=1/2$
and $\et_m = 1$ when $m\neq 0$.

\begin{figure}
  \centering
  \includegraphics[width=\columnwidth]{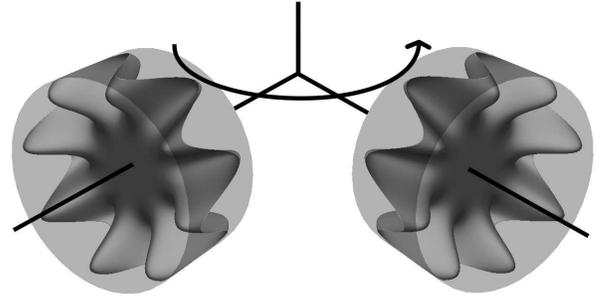}
  \caption{
    Diagram of the sapphire orientation.
    The wave patterns represent
    the dominate axial field
    in the WGH$_{8,0,0}$ mode.
    In this experiment,
    the sapphire radius and height are approximately
    1.5 cm and 3 cm, respectively.
    The oscillator pair is rotated about
    the vertical axis.}
  \label{fig}
\end{figure}

The data used in this analysis are from
an experiment run at the
University of Western Australia
between 2004 and 2006 \cite{cav1}.
The experiment is based on 
two orthogonally aligned
cryogenic sapphire oscillators
lying in the horizontal plane and rotated
around the vertical axis,
as illustrated in Fig.\ \ref{fig}
The resonators were excited at microwave frequencies
to produce a standing-wave whispering-gallery mode
with a dominant electric field in the axial direction and
8 azimuthal field variations (designated WGH$_{8,0,0}$).
The beat frequency was recorded as a function
of time and orientation.
For a more detailed description of the experiment,
see Ref.\ \cite{cav1}.

To calculate the ${\cal M}^{\rm cav}_{njm}$
matrices needed in \eqref{eq:Amm}
the electric fields in the cavities must be modeled.
Maxwell's equations are solved to obtain analytic
approximations for the fields inside and
outside the sapphire \cite{fields}.
For example, the dominate axial component
of the field inside the sapphire crystal
is reasonably described by 
\begin{equation}
  E_{z} \propto J_{8}(695 \text{ m}^{-1} \rho)\, \cos(8\ph)\, \cos(104 \text{ m}^{-1} z) 
\end{equation}
in cylindrical coordinates $(\rh,\ph,z)$.
The other field components are of a similar form.
It is found that over 98\% of
the electric-field energy
is confined within the
actual sapphire crystal.
Consequently,
we neglect the contribution from the
fields outside the sapphire.
Using the fields inside the sapphire,
we generate estimates for the
${\cal M}^{\rm cav}_{njm}$ matrices
via the methods outlined in Ref.\ \cite{km09}.
We then use these to express
the amplitudes in \eqref{eq:CandScomps}
in terms of the coefficients for Lorentz violation.
The results are shown in Table \ref{tab:sens}.

\begin{table*}
  \renewcommand{\tabcolsep}{5pt}
  \renewcommand{\arraystretch}{1.6}
  \begin{tabular}{c|cc|c|c|c|c}
    dimension & $m$ & $m'$ & $C^C_{mm'}$ & $C^S_{mm'}$ & $S^C_{mm'}$ & $S^S_{mm'}$ \\
    \hline\hline
    $d=4$ 
    & 2 & 0 & 
    $-0.27\, \cIdjm{4}{20}$ & 0 & 0 & 0 \\
    & 2 & 1 &
    $0.27\, \re\big(\cIdjm{4}{21}\big)$ & $-0.27\, \im\big(\cIdjm{4}{21}\big)$ &
    $0.52\, \im\big(\cIdjm{4}{21}\big)$ & $0.52\, \re\big(\cIdjm{4}{21}\big)$ \\
    & 2 & 2 &
    $-0.39\, \re\big(\cIdjm{4}{22}\big)$ & $0.39\, \im\big(\cIdjm{4}{22}\big)$ &
    $-0.32\, \im\big(\cIdjm{4}{22}\big)$ & $-0.32\, \re\big(\cIdjm{4}{22}\big)$ \\
    \hline
    $d=6$ 
    & 2 & 0 & 
    $-3.2\, \cfdnjm{6}{220}$ & 0 & 0 & 0 \\
    & 2 & 1 &
    $3.3\, \re\big(\cfdnjm{6}{221}\big)$ & $-3.3\, \im\big(\cfdnjm{6}{221}\big)$ &
    $6.2\, \im\big(\cfdnjm{6}{221}\big)$ & $6.2\, \re\big(\cfdnjm{6}{221}\big)$ \\
    & 2 & 2 &
    $-4.7\, \re\big(\cfdnjm{6}{222}\big)$ & $4.7\, \im\big(\cfdnjm{6}{222}\big)$ &
    $-3.9\, \im\big(\cfdnjm{6}{222}\big)$ & $-3.9\, \re\big(\cfdnjm{6}{222}\big)$ \\
    \hline
    $d=8$
    & 2 & 0 & 
    $0.18\, \cfdnjm{8}{220} $ & 0 & 0 & 0\\[-4pt]
    & & & 
    $-17\, \cfdnjm{8}{420}$ &  & & \\[-4pt]
    & & & 
    $-3.6\, \cfdnjm{8}{440}$ & & & \\
    & 2 & 1 &
    $-0.19\, \re\big(\cfdnjm{8}{221}\big)$ & $0.19\, \im\big(\cfdnjm{8}{221}\big)$ &
    $-0.35\, \im\big(\cfdnjm{8}{221}\big)$ & $-0.35\, \re\big(\cfdnjm{8}{221}\big)$ \\[-4pt]  
    & & & 
    $+18\, \re\big(\cfdnjm{8}{421}\big)$ & $-18\, \im\big(\cfdnjm{8}{421}\big)$ &
    $+33\, \im\big(\cfdnjm{8}{421}\big)$ & $+33\, \re\big(\cfdnjm{8}{421}\big)$ \\[-4pt]
    & & & 
    $-8.6\, \re\big(\cfdnjm{8}{441}\big)$ & $+8.6\, \im\big(\cfdnjm{8}{441}\big)$ &
    $+3.8\, \im\big(\cfdnjm{8}{441}\big)$ & $+3.8\, \re\big(\cfdnjm{8}{441}\big)$ \\
    & 2 & 2 &
    $0.27\, \re\big(\cfdnjm{8}{222}\big)$ & $-0.27\, \im\big(\cfdnjm{8}{222}\big)$ &
    $0.22\, \im\big(\cfdnjm{8}{222}\big)$ & $0.22\, \re\big(\cfdnjm{8}{222}\big)$ \\[-4pt]
    & & & 
    $-25\, \re\big(\cfdnjm{8}{422}\big)$ & $+25\, \im\big(\cfdnjm{8}{422}\big)$ &
    $-21\, \im\big(\cfdnjm{8}{422}\big)$ & $-21\, \re\big(\cfdnjm{8}{422}\big)$ \\[-4pt]
    & & & 
    $+0.84\, \re\big(\cfdnjm{8}{442}\big)$ & $-0.84\, \im\big(\cfdnjm{8}{442}\big)$ &
    $+11\, \im\big(\cfdnjm{8}{442}\big)$ & $+11\, \re\big(\cfdnjm{8}{442}\big)$ \\
    & 2 & 3 &
    $-3.1\, \re\big(\cfdnjm{8}{443}\big)$ & $3.1\, \im\big(\cfdnjm{8}{443}\big)$ &
    $ 1.8\, \im\big(\cfdnjm{8}{443}\big)$ & $1.8\, \re\big(\cfdnjm{8}{443}\big)$ \\
    & 2 & 4 &
    $-8.1\, \re\big(\cfdnjm{8}{444}\big)$ & $ 8.1\, \im\big(\cfdnjm{8}{444}\big)$ &
    $-6.7\, \im\big(\cfdnjm{8}{444}\big)$ & $-6.7\, \re\big(\cfdnjm{8}{444}\big)$ \\
  \end{tabular}
  \caption{\label{tab:sens}
    Nonzero sensitivities to
    coefficients for Lorentz violation
    for dimensions 4, 6, and 8.
    The numbers $m$ and $m'$ give the harmonics
    of the turntable rotation frequency
    and sidereal frequency, respectively.
    The dimension-6 amplitudes are
    in units of $10^{-26}$ GeV$^2$.
    The dimension-8 amplitudes are
    in units of $10^{-52}$ GeV$^4$.}
\end{table*}

Similar to Ref.\ \cite{cav2},
the data is demodulated at twice
the turntable rotation frequency ($2\ph$)
over a period of 500 rotations and
fit to the amplitudes $C_m$ and $S_m$
in \eqref{eq:nubeattwicerot}.
The time derivative of the data
is then weighted, to ensure that
the power spectral density of the
residuals is white, and fit to the
appropriate frequencies of interest
from \eqref{eq:CandScomps}.

To place bounds on the
coefficients for Lorentz violation,
each dimension is considered to
be independent of the others.
For dimensions 4 and 6,
we take combinations of the sine and cosine
terms at each frequency to place
constraints on the real and
imaginary parts of
$\cIdjm{4}{20}$,
$\cIdjm{4}{21}$,
$\cIdjm{4}{22}$,
$\cfdnjm{6}{220}$,
$\cfdnjm{6}{221}$,
and
$\cfdnjm{6}{222}$.
For dimension 8,
the coefficients leading to variations at
three ($m'=3$) and four ($m'=4$)
times the sidereal frequency
can be fully independently constrained.
However,
it is not possible to place
individual bounds on the
dimension-8 coefficients leading
to variations at other
frequencies ($m'=0,1,2$).
From Table \ref{tab:sens},
we see that we are sensitive to
one linear combination of three
coefficients for $m'=0$
and to four combinations of six
real coefficients for $m'=1$ and $m'=2$.
However, we can place constraints
on the nine linear combinations.
The results of our analysis are
given in Table \ref{tab:results}.
We note that several coefficients
are significant at the $1\si$ level.
However, all coefficients are consistent
with zero at the $3\si$ level.

\begin{table*}
  \renewcommand{\tabcolsep}{5pt}
  \renewcommand{\arraystretch}{1.6}
  \begin{tabular}{c|c|r@{$\times$}l}
    dimension & coefficient & \multicolumn{2}{|c}{measurement}  \\
    \hline\hline
    $d=4$ & $\cIdjm{4}{20}$       & $(3\pm 16)$&$ 10^{-15}$\\
    &$\re\big(\cIdjm{4}{21}\big)$ & $(20\pm 23)$&$ 10^{-17}$\\
    &$\im\big(\cIdjm{4}{21}\big)$ & $(137\pm 71)$&$ 10^{-18}$\\
    &$\re\big(\cIdjm{4}{22}\big)$ & $(-4\pm 23)$&$ 10^{-17}$\\
    &$\im\big(\cIdjm{4}{22}\big)$ & $(20\pm 22)$&$ 10^{-18}$\\
    \hline
    $d=6$& $\cfdnjm{6}{220}$       & $(3\pm 13)$&$ 10^{10} \text{ GeV}^{-2}$\\
    &$\re\big(\cfdnjm{6}{221}\big)$& $(17\pm 19)$&$ 10^8 \text{ GeV}^{-2}$\\
    &$\im\big(\cfdnjm{6}{221}\big)$& $(114\pm 59)$&$ 10^7 \text{ GeV}^{-2}$\\
    &$\re\big(\cfdnjm{6}{222}\big)$& $(-3\pm 19)$&$ 10^8 \text{ GeV}^{-2}$\\
    &$\im\big(\cfdnjm{6}{222}\big)$& $(37\pm 40)$&$ 10^7 \text{ GeV}^{-2}$\\
    \hline
    $d=8$& $\cfdnjm{8}{220} - 94 \cfdnjm{8}{420} - 20 \cfdnjm{8}{440}$ 
    &$(5\pm 23)$&$ 10^{37}\text{ GeV}^{-4}$\\
    &$\re\big(\cfdnjm{8}{221}\big) - 98\,\re\big(\cfdnjm{8}{421}\big) + 48\,\re\big(\cfdnjm{8}{441}\big)$
    & $(11\pm 15)$&$ 10^{36}\text{ GeV}^{-4}$\\
    &$\im\big(\cfdnjm{8}{221}\big) - 98\,\im\big(\cfdnjm{8}{421}\big) + 48\,\im\big(\cfdnjm{8}{441}\big)$
    &$(39\pm 15)$&$ 10^{36}\text{ GeV}^{-4}$\\
    &$\re\big(\cfdnjm{8}{221}\big) - 92\,\re\big(\cfdnjm{8}{421}\big) - 11\,\re\big(\cfdnjm{8}{441}\big)$
    & $(58\pm 55)$&$ 10^{35}\text{ GeV}^{-4}$\\
    &$\im\big(\cfdnjm{8}{221}\big) - 92\,\im\big(\cfdnjm{8}{421}\big) - 11\,\im\big(\cfdnjm{8}{441}\big)$
    &$(7\pm 55)$&$ 10^{35}\text{ GeV}^{-4}$\\
    &$\re\big(\cfdnjm{8}{222}\big) - 96\,\re\big(\cfdnjm{8}{422}\big) + 3\,\re\big(\cfdnjm{8}{442}\big)$
    & $(66\pm 29)$&$ 10^{35}\text{ GeV}^{-4}$\\
    &$\im\big(\cfdnjm{8}{222}\big) - 96\,\im\big(\cfdnjm{8}{422}\big) + 3\,\im\big(\cfdnjm{8}{442}\big)$
    &$(-26\pm 28)$&$ 10^{35}\text{ GeV}^{-4}$\\
    &$\re\big(\cfdnjm{8}{222}\big) - 94\,\re\big(\cfdnjm{8}{422}\big) + 49\,\re\big(\cfdnjm{8}{442}\big)$
    & $(-8\pm 24)$&$ 10^{35}\text{ GeV}^{-4}$\\
    &$\im\big(\cfdnjm{8}{222}\big) - 94\,\im\big(\cfdnjm{8}{422}\big) + 49\,\im\big(\cfdnjm{8}{442}\big)$
    &$(-25\pm 25)$&$ 10^{35}\text{ GeV}^{-4}$\\
    &$\re\big(\cfdnjm{8}{443}\big)$& $(-11\pm 15)$&$ 10^{33}\text{ GeV}^{-4}$\\
    &$\im\big(\cfdnjm{8}{443}\big)$& $(-89\pm 53)$&$ 10^{32}\text{ GeV}^{-4}$\\
    &$\re\big(\cfdnjm{8}{444}\big)$& $(-5\pm 29)$&$ 10^{33}\text{ GeV}^{-4}$\\
    &$\re\big(\cfdnjm{8}{444}\big)$& $(-10\pm 28)$&$ 10^{32}\text{ GeV}^{-4}$\\
  \end{tabular}
  \caption{\label{tab:results}
    Measurements of coefficients for Lorentz violation with $1\si$ errors.}
\end{table*}

Our analysis focused on variations
at twice the turntable frequency.
Note that variations at four times the
turntable frequency may also occur, in general.
However, the $\pi/2$ separation of
the resonators results in
an identical shift in each cavity,
and the sensitivities to coefficients at
this frequency cancel out in the beat frequency.
Using a different relative orientation or
introducing a third resonator offset from
the first two by $\pi/4$
would allow the coefficients
$\cfdnjm{8}{440}$,
$\cfdnjm{8}{441}$,
and $\cfdnjm{8}{442}$
to be bound independently.
Different combinations of beat frequencies
between the resonators could then be used
to fully constrain the remaining coefficients.

In summary, 
we have placed constraints
on dimension-6 and dimension-8
camouflage coefficients of the SME.
These represent the first
experimental measurements of
nonrenormalizable Lorentz violation.
We are able to measure five
dimension-6 coefficients and
thirteen combinations of
dimension-8 coefficients.
In our analysis,
we found that the number of higher-dimension
coefficients that can be accessed was restricted
by the use of perpendicular cavities.
Future experiments might achieve sensitivity
to additional combinations of coefficients by
orienting the cavities at other angles or by
introducing a third cavity at an intermediate angle.
Regardless, this work demonstrates the effectiveness
of resonator experiments at detecting
the higher-order Lorentz violation described
by the nonminimal SME.

\begin{acknowledgments}
This work was supported by Australian Research Council Grant No. FL0992016.

\end{acknowledgments}


\begin{thebibliography}{99}

\bibitem{datatables}
{\it Data Tables for Lorentz and CPT Violation,}
V.A.\ Kosteleck\'y and N.\ Russell, arXiv:0801.0287.

\bibitem{ks}
V.A.\ Kosteleck\'y and S.\ Samuel,
Phys.\ Rev.\ D {\bf 39}, 683 (1989);
V.A.\ Kosteleck\'y and R.\ Potting,
Nucl.\ Phys.\ B {\bf 359}, 545 (1991).

\bibitem{km09}
V.A.\ Kosteleck\'y and M.\ Mewes,
Phys.\ Rev.\ D {\bf 80}, 015020 (2009).

\bibitem{cav1}
P.L.\ Stanwix \etal,
Phys.\ Rev.\ D {\bf 74}, 081101 (2006).

\bibitem{cav2}
M.A.\ Hohensee \etal,
Phys.\ Rev.\ D {\bf 82}, 076001 (2010).

\bibitem{micro}
J.\ Lipa \etal,
Phys.\ Rev.\ Lett.\ {\bf 90}, 060403 (2003);
P.\ Wolf \etal,
Gen.\ Rel.\ Grav.\ {\bf 36}, 2352 (2004);
P.\ Wolf \etal,
Phys.\ Rev.\ D {\bf 70}, 051902 (2004);
P.L.\ Stanwix \etal,
Phys.\ Rev.\ Lett.\ {\bf 95}, 040404 (2005);
P.L.\ Stanwix \etal,
Phys.\ Rev.\ D {\bf 74}, 081101 (2006).

\bibitem{optical}
H.\ M\"uller \etal,
Phys.\ Rev.\ Lett.\ {\bf 91}, 020401 (2003);
S.\ Herrmann \etal,
Phys.\ Rev.\ Lett.\ {\bf 95}, 150401 (2005);
P.\ Antonini \etal,
Phys.\ Rev.\ A {\bf 71}, 050101 (2005);
{\bf 72}, 066102 (2005);
M.E.\ Tobar, P.\ Wolf, and P.L.\ Stanwix,
Phys.\ Rev.\ A {\bf 72}, 066101 (2005);
S.\ Herrmann \etal,
Phys.\ Rev.\ D {\bf 80}, 105011 (2009);
Ch.\ Eisele, A.Yu.\ Nevsky, and S.\ Schiller,
Phys.\ Rev.\ Lett.\ {\bf 103}, 090401 (2009).

\bibitem{cav3}
H.\ M\"uller \etal,
Phys.\ Rev.\ D {\bf  67}, 056006 (2003); 
H.\ M\"uller \etal,
Phys.\ Rev.\ D {\bf  68}, 116006 (2003);
M.E.\ Tobar \etal,
Phys.\ Rev.\ D {\bf 71}, 025004 (2005);
H.\ M\"uller, 
Phys.\ Rev.\ D {\bf  71}, 045004 (2005);
H.\ M\"uller \etal,
Phys.\ Rev.\ Lett.\ {\bf 99}, 050401 (2007);
M.\ Mewes,
Phys.\ Rev.\ D {\bf 78}, 096008 (2008);
M.E.\ Tobar \etal,
Phys.\ Rev.\ D {\bf 80}, 125024 (2009).

\bibitem{mm}
A.A.\ Michelson and E.W.\ Morley,
Am.\ J.\ Sci.\ {\bf 34}, 333 (1887);
Phil.\ Mag.\ {\bf 24}, 449 (1887).

\bibitem{ck}
D.\ Colladay and V.A.\ Kosteleck\'y, 
Phys.\ Rev.\ D {\bf 55}, 6760 (1997);
{\bf 58}, 116002 (1998).

\bibitem{akgrav}
V.A.\ Kosteleck\'y,
Phys.\ Rev.\ D {\bf 69}, 105009 (2004).

\bibitem{bire}
S.M.\ Carroll, G.B.\ Field, and R.\ Jackiw,
Phys.\ Rev.\ D {\bf 41}, 1231 (1990);
V.A.\ Kosteleck\'y and M.\ Mewes,
Phys.\ Rev.\ Lett.\ {\bf 87}, 251304 (2001);
{\bf 97}, 140401 (2006);
{\bf 99}, 011601 (2007);
Q.\ Exirifard, arXiv:1010.2054.

\bibitem{km_apjl}
V.A.\ Kosteleck\'y and M.\ Mewes,
Ap.\ J.\ Lett.\ {\bf 689}, L1 (2008).

\bibitem{fermi}
V.\ Vasileiou, arXiv:1008.2913.

\bibitem{km02}
V.A.\ Kosteleck\'y and M.\ Mewes,
Phys.\ Rev.\ D {\bf 66}, 056005 (2002).

\bibitem{fields}
M.\ Tobar and A.\ Mann,
IEEE Trans.\ on MTT {\bf 39}, 2077 (1991);
M.\ Tobar \etal,
Lect.\ Notes Phys.\ {\bf 702}, 416 (2006).



\end{thebibliography}
\end{document}